\documentclass[twocolumn,showpacs,preprintnumbers,amsmath,amssymb,superscriptaddress]{revtex4}

\usepackage{graphicx}
\usepackage{graphicx}
\begin{document}
\title{Effects of chemical pressure on the Fermi surface and band dispersion in the electron-doped high-$T_{\rm c}$ superconductors}

\author{M. Ikeda}
\affiliation{Department of Physics and Department of Complexity 
Science and Engineering, University of Tokyo, 
Hongo 7-3-1, Bunkyo-ku, Tokyo 113-0033, Japan}

\author{T. Yoshida}
\affiliation{Department of Physics and Department of Complexity 
Science and Engineering, University of Tokyo, 
Hongo 7-3-1, Bunkyo-ku, Tokyo 113-0033, Japan}

\author{A. Fujimori}
\affiliation{Department of Physics and Department of Complexity 
Science and Engineering, University of Tokyo, 
Hongo 7-3-1, Bunkyo-ku, Tokyo 113-0033, Japan}

\author{M. Kubota}
\affiliation{Institute of Material Structures Science, High Energy Accelerator Research Organization (KEK), 
Oho 1-1, Tsukuba, Ibaraki 305-0801, Japan}

\author{K. Ono}
\affiliation{Institute of Material Structures Science, High Energy Accelerator Research Organization (KEK), 
Oho 1-1, Tsukuba, Ibaraki 305-0801, Japan}

\author{Hena Das}
\affiliation{S. N. Bose National Centre for Basic Science, JD Block, Sector 3, Salt Lake City, Kolkata 700098, 
India}

\author{T. Saha-Dasgupta}
\affiliation{S. N. Bose National Centre for Basic Science, JD Block, Sector 3, Salt Lake City, Kolkata 700098, 
India}

\author{K. Unozawa}
\affiliation{Department of Advanced Materials Science, University of Tokyo, 
Kashiwanoha 5-1-5, Kashiwa, Chiba 277-8561, Japan}

\author{Y. Kaga}
\affiliation{Department of Advanced Materials Science, University of Tokyo, 
Kashiwanoha 5-1-5, Kashiwa, Chiba 277-8561, Japan}

\author{T. Sasagawa}
\affiliation{Department of Advanced Materials Science, University of Tokyo, 
Kashiwanoha 5-1-5, Kashiwa, Chiba 277-8561, Japan}
\affiliation{Materials and Structures Laboratory, Tokyo Institute of Technology, 
Nagatsuta 4259, Midori-ku, Yokohama, Kanagawa 226-8503, Japan}

\author{H. Takagi}
\affiliation{Department of Advanced Materials Science, University of Tokyo, 
Kashiwanoha 5-1-5, Kashiwa, Chiba 277-8561, Japan}

\date{\today}
\begin{abstract}

We have performed angle-resolved photoemission spectroscopy (ARPES) measurements and first-principles electronic structure calculations on the electron-doped high-$T_{\rm c}$ superconductors $Ln_{1.85}$Ce$_{0.15}$CuO$_{4}$ ($Ln$ = Nd, Sm, and Eu). The observed Fermi surface and band dispersion show such changes that with decreasing ionic size of $Ln^{3+}$, the curvature of the Fermi surface or $-t'/t$ decreases, where $t$ and $t'$ are transfer integrals between the nearest-neighbor and next-nearest-neighbor Cu sites, respectively. The increase of $t$ with chemical pressure is found to be significant, which may explain the apparently inconsistent behavior seen in the hole-doped La$_{2-x}$Sr$_{x}$CuO$_{4}$ under epitaxial strain [M. Abrecht $et$ $al$., Phys. Rev. Lett. ${\bf 91}$, 057002 (2003)]. A gap due to the antiferromagnetism opens even in the nodal region for the Sm and Eu compounds, and the gap size increases in going from $Ln$ = Sm to Eu.

\end{abstract}
\maketitle

Since the discovery of the high-$T_{\rm c}$ superconductors (HTSCs) \cite{highTc}, a large number of studies have been performed in order to obtain higher critical temperatures. Among them, pressure effects have attracted much attention because it causes a dramatic increase of $T_{\rm c}$ in many systems. For example, in HgBa$_{2}$Ca$_{2}$Cu$_{3}$O$_{4}$, which has the highest $T_{\rm c}$ among HTSCs, $T_{\rm c}$ rises from 135 K to 164 K in a hydrostatic pressure \cite{Gao1994}. In addition to mechanical pressure \cite{Chen1991, Gugenberger1994, Nohara1995, Meingast1996, Nakamura2000}, the effect of epitaxial strain in thin films grown on single crystalline substrates \cite{Locquet1996, Sato1997, Locquet1998}, and the effect of ``chemical pressure", where the lattice constants are varied through substitution of ions with different ionic radii \cite{Markert1990, Uzumaki1991, Naito2000}, have been studied so far. The effects of epitaxial strain and ``chemical pressure'' on $T_{\rm c}$ are consistent with the mechanical pressure in many cases \cite{Chen1991, Gugenberger1994, Nohara1995, Meingast1996, Locquet1996, Sato1997, Locquet1998, Nakamura2000}, but the mechanism of the $T_{\rm c}$ changes has not been understood yet. Also, differences in $T_{\rm c}$ between different cuprate families are expected to provide crucial information about the mechanism of high-$T_{\rm c}$ superconductivity and have been discussed extensively \cite{Pavarini2001, Tanaka2004}. According to first-principles calculations, the shape of the Fermi surface strongly depends on the distance $d_{\rm Cu-Oap}$ between the copper and the apical oxygen. A long $d_{\rm Cu-Oap}$ leads to a strong curvature of the Fermi surface, namely, a large value of $-t'/t$ \cite{Pavarini2001}, where $t$ and $t'$ denote transfer integrals between the nearest-neighbor and next-nearest-neighbor Cu sites, respectively, in the single-band tight-binding model. However, the angle-resolved photoemission spectroscopy (ARPES) results of La$_{2-x}$Sr$_{x}$CuO$_{4}$ (LSCO) under epitaxial strain have demonstrated that compressed in-plane lattice constant and hence increased $d_{\rm Cu-Oap}$ resulted in a decrease of $-t'/t$ \cite{Abrecht2003, Cloetta2006}, contrary to the material dependence of $-t'/t$ \cite{Pavarini2001}. That is, the relationship between the crystal structure and the electronic structure of cuprates appears to be more complicated.

In order to clarify the above issues, we have performed ARPES measurements and first-principles electronic structure calculations on the electron-doped HTSCs $Ln_{2-x}$Ce$_{x}$CuO$_{4}$ ($Ln$ = Nd, Sm, and Eu), where with decreasing ionic radius of $Ln^{3+}$ from $Ln=$ Nd to Eu, ``chemical pressure" increases, that is, both in-plane and out-of-plane lattice constants become small \cite{Markert1990, Uzumaki1991} and $T_{\rm c}$ decreases \cite{Markert1990, Naito2000}. Since these materials have no apical oxygen, the shape of the Fermi surface is not related to the $d_{\rm Cu-Oap}$, but determined by other contributions such as the in-plane lattice constant. Furthermore, no empirical relationship between $-t'/t$ and $T_{\rm c}$ has been known for the electron-doped cuprates. It is therefore highly desired to investigate the electronic structure of the $Ln_{2-x}$Ce$_{x}$CuO$_{4}$ by ARPES. The present results show that chemical pressure reduces the curvature of the Fermi surface as in the case of the ARPES results on LSCO thin films \cite{Abrecht2003}. Further analysis of the band dispersions has revealed that the increase of $t$ as a result of the decrease of the in-plane lattice constant has a significant effect on the Fermi surface shape.

High-quality single crystals of optimally doped Nd$_{1.85}$Ce$_{0.15}$CuO$_{4}$ (NCCO), Sm$_{1.85}$Ce$_{0.15}$CuO$_{4}$ (SCCO) and Eu$_{1.85}$Ce$_{0.15}$CuO$_{4}$ (ECCO) were grown by the traveling solvent floating zone method. The $T_{\rm c}$'s of NCCO, SCCO, and ECCO were $\sim$22 K, $\sim$16 K, and 0 K, respectively. There was slight deviation in the Ce content and/or the oxygen stoichiometry from the indicated composition, as reflected on small differences in the Fermi surface areas among NCCO, SCCO, and ECCO, as we shall see below. The ARPES measurements were performed at beamline 28A of Photon Factory (PF), Institute of Materials Structure Science, High Energy Accelerators Research Organization (KEK), using incident photons with energy of 55 eV. We used a SCIENTA SES-2002 electron-energy analyzer in the angle mode, and used a five-axis manipulator \cite{Aiura2003}. The total energy resolution and momentum resolution were 15 meV and 0.01$\pi$, respectively. Samples were cleaved {\it in situ} under an ultrahigh vacuum of 10$^{-11}$ Torr to obtain clean surfaces, and were measured at $\sim$ 10 K. The Fermi edge of gold was used to determine the Fermi level ($E_{\rm F}$) position and the instrumental resolution before and after the ARPES measurements.

Figure 1 shows plots of ARPES intensity at $E_{\rm F}$ in NCCO, SCCO, and ECCO in two-dimensional momentum space. The suppression of the intensity is seen near the ``hot spots", i.e., the intersecting points of the paramagnetic Fermi surface and the antiferromagnetic Brillouin zone boundary [see the inset in Fig. 1(c)]. This suppression is due to the scattering of electrons at $E_{\rm F}$ by antiferromagnetic fluctuations or by (quasi)static antiferromagnetic correlation with wave vector ($\pi,\pi$) as discussed in previous ARPES studies \cite{NP2, Matsui1}. We present energy distribution curves (EDCs) in the nodal direction in Fig. 2, and ARPES intensity plot in $E-k$ space in the nodal direction in Fig. 3(a), (b), and (c). In Fig. 2, the gap or leading edge (LE) shift is observed for SCCO (LE shift: $\sim$ 5 meV) and ECCO (LE shift: $\sim$ 30 meV) but not for NCCO, consistent with a recent ARPES study on Sm$_{1.86}$Ce$_{0.14}$CuO$_{4}$ \cite{Park2007}.

\begin{figure}
\begin{center}
\includegraphics[width=8.5cm]{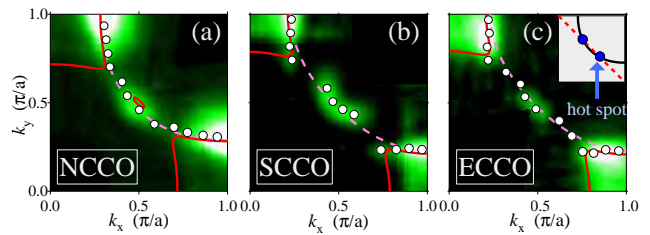}
\caption{(Color online) ARPES intensity within $\pm30$ meV of the Fermi level ($E_{\rm F}$) plotted in momentum space for Nd$_{1.85}$Ce$_{0.15}$CuO$_{4}$ (NCCO) (a), Sm$_{1.85}$Ce$_{0.15}$CuO$_{4}$ (SCCO) (b), and Eu$_{1.85}$Ce$_{0.15}$CuO$_{4}$ (ECCO) (c). The data were taken over a Brillouin zone octant and symmetrized with respect to the (0,0)-$(\pi, \pi)$ line. White circles show the peak positions of momentum distribution curves (MDCs) at $E_{\rm F}$, indicating the shape of the Fermi surface or underlying Fermi surface. Solid red curves and dashed pink curves show the Fermi surface obtained by tight-binding fit to the ARPES data assuming the paramagnetic and antiferromagnetic band structures, respectively. Inset: Schematic diagram of the hot spot. Black curve and red dashed line represent the Fermi surface and the antiferromagnetic Brillouin zone boundary, respectively.}
\end{center}
\end{figure}

White circles in Fig. 1 represent the peak positions of the momentum distribution curves (MDCs) at $E_{\rm F}$ and represent the Fermi surface or remnant Fermi surface. As seen from Fig. 1, the curvature of the Fermi surface in NCCO is the strongest among the three materials. We quantitatively evaluated the difference of the curvature among the three compounds using a tight-binding model as follows. We used two-dimensional antiferromagnetic tight-binding model:

\begin{eqnarray}
\epsilon - \mu= \varepsilon_{0} \pm \sqrt{\Delta E^{2}+4t^{2}(\cos{k_{x}a}+\cos{k_{y}a})^{2}}\nonumber \\-4t'\cos{k_{x}a} \cos{k_{y}a}-2t''(\cos{2k_{x}a}+\cos{2k_{y}a}) \nonumber,
\end{eqnarray}
where $t$, $t'$, and $t''$ are transfer integrals between the nearest-neighbor, second-nearest-neighbor, and third-nearest-neighbor Cu sites, respectively, $\varepsilon_{0}$ represents the center of the band relative to the chemical potential $\mu$, and 2$\Delta E$ is the potential energy difference between the spin-up and spin-down sublattices.

First, we fitted the calculated Fermi surface to the set of white circles in Fig. 1 by adjusting the parameters $-t'/t$ and $-\varepsilon_{0}/t$. We note that the curvature of the Fermi surface depends on the ratio $-t'/t$ and not on each of $t$ and $t'$, and only very weakly on $-\varepsilon_{0}/t$. Assuming that $-t''/t'= 0.50$, we obtained $-t'/t=$ 0.20, 0.12, and 0.11 for NCCO, SCCO, and ECCO, respectively, and $\varepsilon_{0}/t=$ -0.12, 0, and -0.05 for NCCO, SCCO, and ECCO, respectively. The Fermi surface area was allowed to deviate slightly from what is expected for 15 $\%$ Ce doping because of non-stoichiometry. Note that when we performed the same analysis retaining only the second-nearest neighbor Cu hopping ($t''=0$), we obtained $-t'/t=$ 0.40, 0.23, and 0.21 for NCCO, SCCO, and ECCO, respectively, showing a similar tendency to the case of $-t''/t'=0.50$. The present fitted results demonstrate that the curvature of the Fermi surface monotonically decreases from NCCO to SCCO to ECCO.

\begin{figure}
\begin{center}
\includegraphics[width=8.5cm]{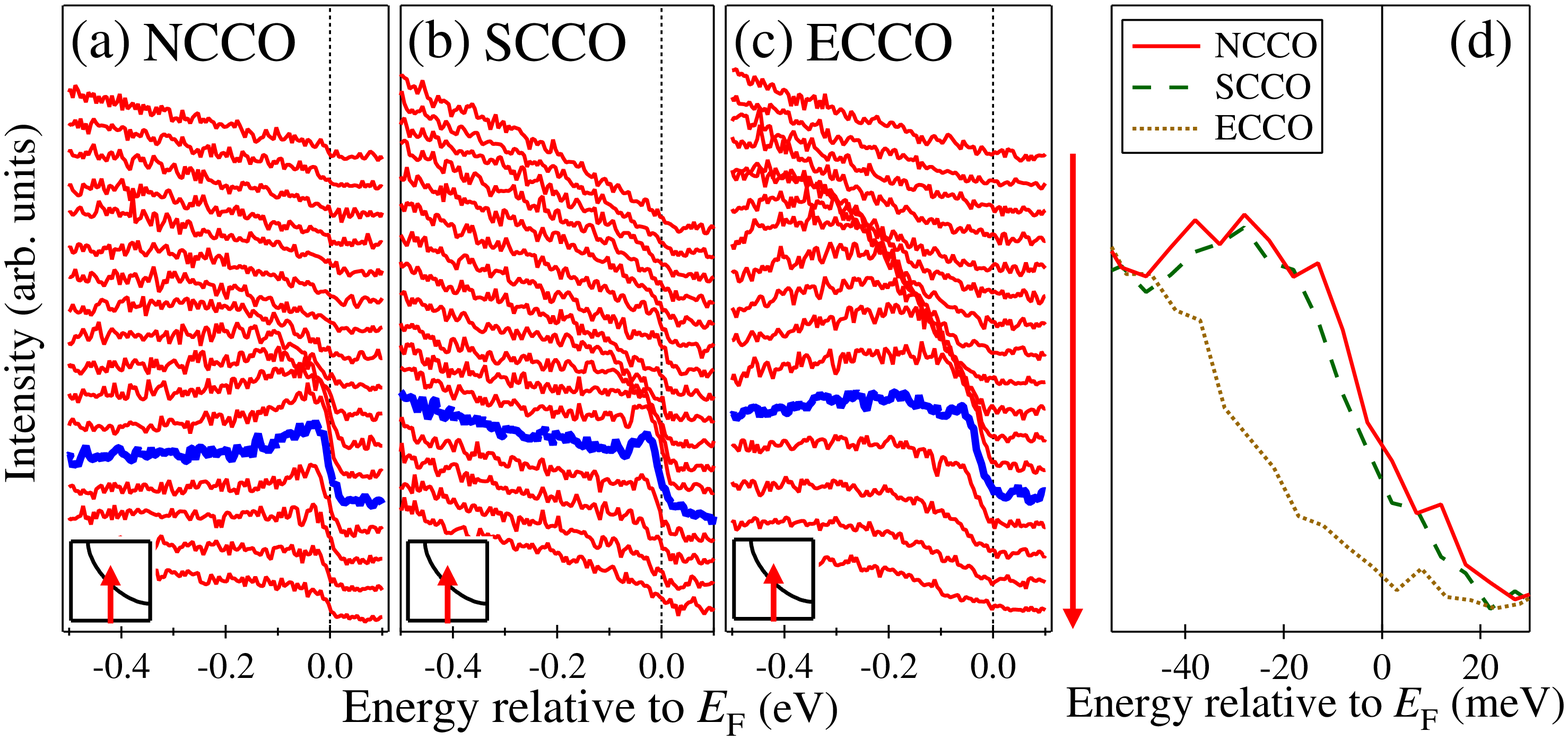}
\caption{(Color online) Energy distribution curves (EDCs) for NCCO (a), SCCO (b), and ECCO (c) around the nodal $k_{\rm F}$ point. Blue thick EDCs represent EDCs at $k_{\rm F}$. The direction of the cut is shown in the insets. (d) EDCs of NCCO, SCCO and ECCO at the nodal $k_{\rm F}$ point.}
\end{center}
\end{figure}

Next, we evaluated the absolute values of $t$ and $t'$ by fitting the calculated band dispersion to the MDC and/or EDC peak positions with $-t'/t$ and $-\varepsilon_{0}/t$ fixed at the above values. Because the position of the ``flat band'' at $\sim(\pi,0)$ is sensitive to $-t'$, it can be used to determine $t'$ and $t$ separately. Figure 3 shows the $E-k$ plot of ARPES intensity for NCCO, SCCO, and ECCO along two different cuts as shown in the inset. Panels (d)-(f) show that the positions of the band around the $(\pi, 0)$ point becomes shallower in going from NCCO to ECCO, indicating that $-t'$ decreases. For the plot of the antiferromagnetic tight-binding bands, we have assumed that the spectral intensity denoted by the size of the point is proportional to the projected weight of the antiferromagnetic band to the paramagnetic band. In the nodal region [panels (a)-(c)] the experimental band dispersion as well as spectral weight is well explained by the antiferromagnetic band while around the $(\pi, 0)$ region [panels (d)-(f)], the experimental one cannot be explained by the antiferromagnetic band well but rather by the paramagnetic band ($\Delta E=0$). These fitting results may indicate that the $k$-dependence of the antiferromagnetic gap $\Delta E$ exists beyond standard band theory of the antiferromagnetic state, as predicted by a recent variational Monte-Carlo study \cite{Chou2007}.

In order to see the effect of ``chemical pressure'' on the band structure from first-principles calculation, we have carried out muffin-tin-orbital (MTO) based NMTO calculations \cite{Andersen2000} within the framework of local density approximation (LDA) and extracted the transfer integrals corresponding to $t$, $t'$, and $t''$ in the tight-binding model. For this purpose, an effective Cu $d_{x^{2}-y^{2}}$ basis was defined by means of the downfolding procedure, by integrating out all the degrees of freedom related to $Ln$, O and Cu except for Cu $d_{x^{2}-y^{2}}$. The effective basis, constructed in this manner, serves the purpose of the Wannier-like function corresponding to the single band crossing the $E_{\rm F}$. The real-space Hamiltonian defined in the basis of these effective, Wannier-like orbitals provides the information about the various transfer integrals, connecting various Cu sites. This method has been applied successfully in the case of hole-doped cuprate compounds \cite{Pavarini2001}. Calculations have been carried out for $Ln=$Nd and Sm. The calculated $-t'/t$ turned out to be 0.34 ($t$ = 0.41 eV, $t'$ = -0.14 eV) for NCCO and 0.29 ($t$ = 0.44 eV, $t'$ = -0.13 eV) for SCCO. Such moderate change in $-t'/t$, compared with the hole doped cuprate family is expected considering the absence of apical oxygen in the $T'$ structure, one of the key controlling factor in determining $-t'/t$ \cite{Pavarini2001}. Though the quantitative estimates of $-t'/t$ are somewhat different from those obtained from experimental data, the trend is correctly described as shown in Fig.4. According to cellular dynamical mean-field theory \cite{Civelli2005}, $-t'/t$ is reduced by electron correlation over the LDA estimate for electron-doped cuprates and increased for hole-doped systems, with the effect being strong for electron-doped case at low doping level. This may explain the disparity in the quantitative values of $-t'/t$ as seen in Fig. 4, where for LSCO a good agreement in $-t'/t$ between the LDA and the experimental estimate was observed.

\begin{figure}
\begin{center}
\includegraphics[width=8.5cm]{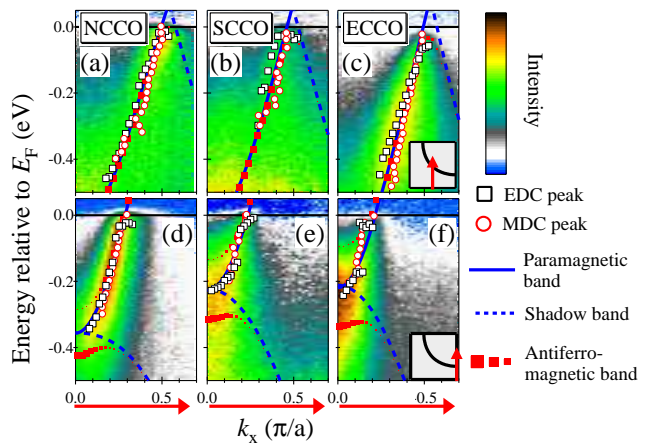}
\caption{(Color online) ARPES intensity plot in energy-momentum space in the nodal and antinodal regions of the Brillouin zone compared with tight-binding energy bands. (a), (d): NCCO, (b), (e): SCCO, and (c), (f): ECCO. The cuts are shown in the inset. The spectral intensity of the antiferromagnetic band is represented by the size of squares.}
\end{center}
\end{figure}

Considering the nearest neighbor Cu-Cu transfer integral $t$ to be given by the relationship $t = 2 (t_{pd})^{2}/(\varepsilon _{d}-\varepsilon _{p})$, where $t_{pd}$ and $\varepsilon _{d}-\varepsilon _{p}$ are the Cu 3$d$ - O 2$p$ transfer integral and the onsite energy differences, respectively, defined within the three band model, the increase of $t$ in going from Nd to Sm is contributed both by the increase of $t_{pd}$ due to contraction of the lattice, as well as due to decrease of $(\varepsilon _{d}-\varepsilon _{p})$ caused by the replacement of Nd by Sm. Our NMTO-downfolding calculations keeping oxygen-$p \sigma$ degrees of freedom active in addition to Cu $d_{x^{2}-y^{2}}$ gives a 8$\%$ change in $t_{pd}$, while the rest is contributed by the change in $(\varepsilon _{d}-\varepsilon _{p})$. The decrease of the second nearest neighbor hopping, $t'$ in moving from Nd to Sm is, on the other hand, contributed by the decrease in the O-O hopping, $t_{pp}$ as obtained in our three band calculations, presumably caused by the different cation covalency effect between Nd and Sm.

We now turn our attention to the case of strained LSCO thin films. In LSCO, the reduced in-plane lattice constant by compressive strain lead to the decrease of $-t'/t$ \cite{Abrecht2003}, as in the present electron-doped system. Hence, the experimental results indicate that the reduced in-plane lattice constant resulted in the reduced value of $-t'/t$ both under chemical pressure and epitaxial strain and both for the hole-doped and electron-doped compounds. Also, since the electron-doped HTSCs have no apical oxygen, the shape of the Fermi surface in our experiment is not related to $d_{\rm Cu-Oap}$ but is related to the in-plane lattice constant including the effect of O-O hopping. Therefore, we consider that in the strained LSCO film the effects of in-plane lattice constant is stronger than those of $d_{\rm Cu-Oap}$.

\begin{figure}
\begin{center}
\includegraphics[width=8cm]{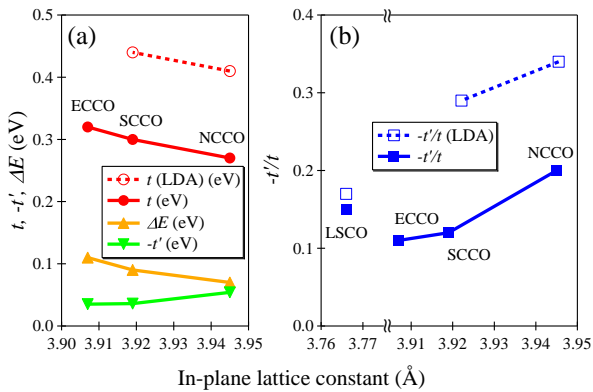}
\caption{(Color online) (a), (b): Tight-binding parameters as functions of in-plane lattice constant. For definition, see text. As the in-plane lattice constant decreases, $-t'$ and $-t'/t$ decreases while $t$ and $\Delta E$ increase. Dashed lines are results from the LDA calculations. LSCO data are taken from Refs. \cite{Pavarini2001, Yoshida2006}.}
\end{center}
\end{figure}

Finally, we discuss the relationship between $-t'/t$ values and the superconductivity. In the hole-doped HTSCs, there is an empirical trend that the larger $-t'/t$ is, the higher $T_{\rm c}$ becomes \cite{Pavarini2001}. The present results on $-t'/t$ in the electron-doped system follow the same trend, but considering LDA estimates of $-t'/t$, the predicted $T_{\rm c}$ would be higher, $\sim90$ K and $\sim70$ K for NCCO and SCCO, respectively. If we employ the experimental values for $-t'/t$, instead, then $T_{\rm c}$ would be $\sim40$ K, 25 K for NCCO and SCCO, in somewhat better agreement with the empirical trend, but still overestimating the difference in $T_{\rm c}$ between NCCO and SCCO. The gap opening in the nodal direction may explain the generally low $T_{\rm c}$'s in the electron-doped systems compared with hole-doped ones as well as the systematic suppression of $T_{\rm c}$ with chemical pressure because electrons near the nodal point significantly contribute to the superconductivity, according to the Raman scattering studies \cite{Qazilbash2005} and variational cluster calculations \cite{Aichhorn2006}. In addition, in going from NCCO to ECCO, the compression of the out-of-plane lattice constant becomes strong compared with that of the $a$-axis \cite{Markert1990}. The decrease of the inter-layer distance may cause the development of three-dimensional antiferromagnetic order, leading to the large $\Delta E$. Further studies are necessary to clarify whether the effect of variation in $-t'/t$ or that in $\Delta E$ is more important for the change of $T_{\rm c}$.

In conclusion, we have performed ARPES and first-principles electronic structure calculation studies of NCCO, SCCO, and ECCO in order to elucidate the variation of electronic structure by ``chemical pressure". As the in-plane lattice constant decreases, $-t'/t$ decreases and $t$ increases, consistent with the previous ARPES results on strained LSCO films. This suggests that the variation of the in-plane lattice constant has a great influence on the electronic structures. We consider that the decrease of $T_{\rm c}$ with increasing ``chemical pressure'' is attributed to the change in the Fermi surface and/or the gap opening around the nodal point.

We are grateful to T. Tohyama, T. K. Lee, C. M. Ho, C. P. Chou, and N. Bontemps for enlightening discussion. This work was done under the approval of the Photon Factory Program Advisory Committee (Proposal No. 2006S2-001) and was supported by a Grant-in-Aid for Scientific Research in Priority Area ``Invention of Anomalous Quantum Materials'' from the Ministry of Education, Culture, Sports, Science and Technology, Japan. We also thank the Material Design and Characterization Laboratory, Institute for Solid State Physics, University of Tokyo, for the use of the SQUID magnetometer.

\end{document}